# Design, Fabrication and Testing of a D-Shaped High Temperature Superconducting Magnet

Upendra Prasad, Mahesh Ghate, Piyush Raj, Deven Kanabar, Pankaj Varmora, Swati Roy, Arun Panchal, Dhaval Bhavsar, Anees Bano, Nitish Kumar, Bhadresh Parghi, Akhilesh Yadav, Mohd.Umer, Vijay Vasava, Raton Mandal, Rajkumar Ahirwar, Megha Thaker

*Institute for Plasma Research, Bhat, Gandhinagar-382428, Gujarat, India*

*Abstract*- **High-temperature technical superconductors are potential candidates for compact and high-field tokamak magnets. The demand for higher fusion power can be met with an on-axis high magnetic field due to toroidal magnets. An R&D activity has been initiated at the Institute for Plasma Research, India, to develop a compact D-shaped superconducting magnet utilizing REBCO high-temperature superconducting tapes. Under this initiative, a toroidal configuration with a major radius of 0.42 m, consisting of eight D-shaped, four poloidal field, and a central solenoid high-temperature superconducting magnets producing an on-axis toroidal magnetic field of 0.23 T has been conceptualized. The fabrication feasibility of a D-shaped coil for this toroidal configuration also envisaged using stacked high-temperature superconducting cable. In this paper, we report the design of a compact D-shaped coil, the fabrication of a long length HTS cable, a winding pack, and its integration with a cryogenic casing and vacuum enclosure. The winding pack terminations, joints, its interfacing with the power supply, and performance testing are also reported in this paper.**

## I. INTRODUCTION

The present commercially available REBCO high temperature superconducting tapes (HTS) are capable of carrying higher transport current at a reasonably higher temperature in comparison to NbTi and $Nb_3Sn$ low temperature superconductors (LTS). This advantage could be utilize to significantly reduce the size of the HTS coil winding pack, while maintaining the power consumption level similar to that of copper coils. These advantages of HTS over LTS and Copper are technical solutions for developing compact and high current density coils for magnetic fusion. HTS can be used to meet the demand of the high magnetic field in the optimum available space for coils. Multiple designs of toroidal field (TF) magnets [1], [2], [3], [4] are evolving to produce a higher toroidal magnetic field for magnetic confinement of plasma. Prototype TF coil winding packs [5], [6], [7] are being realized worldwide to test the technical viability of HTS conductor for fusion magnets application. Recently, an HTS TF coil [8] was fabricated and tested up to 20 T at 20 K, surpassing the limit of the LTS TF magnets [9]. India has also worked out a fusion road map [10] to conceive a high temperature superconductor based magnetic fusion machine. An R&D program has also been initiated to develop the HTS magnet enabling technologies in line with the Indian fusion road map at the Institute for Plasma Research, Gandhinagar, Gujarat. REBCO magnets [11], [12] and high current cables [13], [14] are being developed under this R&D initiative.

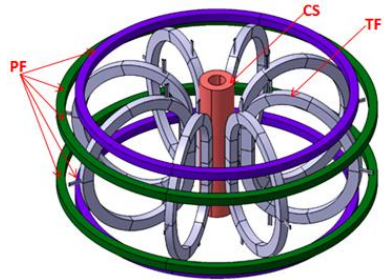

Fig. 1. Magnetic configuration of HTS coils

The schematic of a small-scale toroidal magnetic configuration with major and minor radii of 0.42 m and 0.33 m is conceptualized and shown in Fig. 1. It consists of a central solenoid (CS), eight D-shaped toroidal field (TF) coils, and four poloidal field (PF) coils for plasma initiation, confinement, and shaping. As a step forward, the design, fabrication, and testing of a stacked HTS cable based D-shaped coil with a height of 1.1 m and a 0.7 m width, has been undertaken to identify and asses the technical challenges involved in the fabrication. The design of a D-shaped coil and HTS cable is discussed in sections II and III of this paper. The fabrication of the coil, including the winding pack, cable terminations & joints, integration in casing, and sensors & instrumentation, is described in section IV. The experimental test setup, results & discussions and summary are elaborated in sections V, VI and VII, respectively.

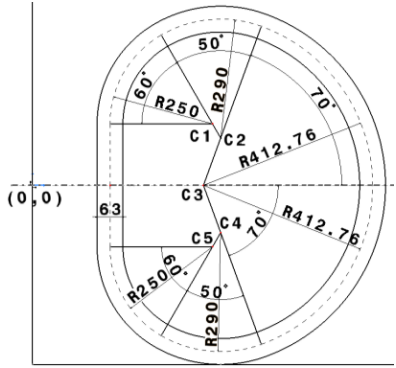

Fig.2. Major dimensions for small-scale toroidal magnetic configuration

## II. DESIGN OF D-SHAPED COIL

The design of the D-shaped coil is based on a bending-free and constant tension profile for the tokamak [15], [16]. Figure.2 shows the major dimensions of a D-shaped coil for the toroidal magnetic configuration as discussed in the previous section. The coil geometry has centers, C1 (439,150), C2 (459,115), C3 (0,0), C4(-459,-115), and C5(-439,-150) with arc of radii of 250 mm, 290 mm, and 412.76 mm, and a straight section of length 150 mm, respectively. The height-to-width ratio of the coil is approximately 1.25. The distances of the inner and outer edges of the D-shaped coil from the central axis of magnetic configuration are around 0.189 m and 0.830 m. The radial width of the D-shaped coil is 63 mm to accommodate conductor, insulation, and coolant. The simulation for the estimation of the toroidal magnetic field profile for the toroidal configuration of eight D-shaped coils was carried out using ANSYS Maxwell. The magnetic field profile at the coil and between two coils is shown in Fig.3. The magnetic field of 0.231 T was estimated at the major radius of 0.42 m from the center of the toroidal assembly with an operating current per turn of 2.5 kA. The peak magnetic field at the straight leg of the D-shaped coil winding pack is around 0.67 T, and I × B transverse Lorentz load of 1.62 kN/m. To realize the associated technological challenges for the fabrication of a D-shaped HTS coil, a prototype coil (PC) with a double pancake consisting of 13 turns, conductor size of 13.5 mm × 9.5 mm, and operating current per turn of ~1 kA at 77 K, a self-field has been realized. The major parameters for the designed D-shaped coil for the toroidal configuration and prototype coil to be fabricated are summarized in Table 1.

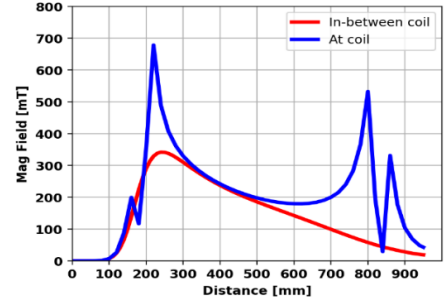

Fig.3. Magnetic Field profile at coil and between two coils f

TABLE 1: PARAMETERS OF D-SHAPED COILS

| Parameters | Designed Coil | Prototype Coil |
|---|---|---|
| Winding Pack (WP) Height (mm) | 810.72 | 810.72 |
| WP Width (mm) | 640.78 | 640.78 |
| Height to width Ratio | 1.265 | 1.265 |
| No of Arc in half | 3 | 3 |
| WP Minimum turn radius (mm) | 250 | 250 |
| Distance between two opposite coil (mm) | 300 | - |
| Number of Turns/coil | 24 | 13 |
| Operating Current (A) | 2500 | 1220 |
| No of Coils | 8 | 1 |
| No of pancakes | 2 | 2 |
| Coil perimeter | 2.34 | 2.34 |
| Total conductor length for all the coils (m) | 561.2 | 30.42 |
| Required length of HTS tape (m) | ~3367 | ~183 |
| Coolant | $LN_2$ | $LN_2$ |
| Operating temperature (K) | 77 | 77 |
| Estimated inductance (µH) | 845 | 248 |
| Peak field at inner leg (T) | 0.678 | 0.115 |

## III. DESIGN OF HTS CABLE

The zero-dimensional (0-D) approach has been used for the basic design of the HTS cable. The design of cable is governed by stability criteria, critical current, critical temperature, critical field, copper to superconducting ratio, and energy margin. The assumptions for this approach were already discussed [17]. The analytical equations used to estimate the upper critical current, Stekly Parameter, and the lower critical current for the design of this cable are mentioned below in (1), (2), and (3).

$$(I_{lim})^{upp} = [\{A_{Cu} \cdot P_w \cdot h \cdot (T_c - T_{op})\}/\rho_{Cu}]^{1/2} \qquad (1)$$

$$\alpha = \{I_{op}/(I_{lim})^{upp}\}^2 \qquad (2)$$

$$(I_{lim})^{low} = [A_{Cu} \cdot P_w \cdot h \cdot (T_c - T_{op})/I_c \cdot \rho_{Cu}] \qquad (3)$$

In (1), $I_{lim}$, $A_{cu}$, $P_w$, $h$, $T_c$, $T_{op}$, and $\rho_{Cu}$ are limiting current, stabilizer copper cross-sectional area, cooled perimeter, heat transfer coefficient, critical temperature, operating current, and copper resistivity. In (2), $\alpha$, $I_{op}$ is the Stekly stability parameter and operating current. In (3), $I_c$ is the critical current of the conductor. The major estimated parameters of the HTS cable are summarized in Table 2.

TABLE 2: PARAMETERS FOR HTS CABLE

| Cable Parameters | Values |
| --- | --- |
| HTS tape width / thickness | 12.1 mm / 0.14 mm |
| Cu tape width / thickness | 12 mm / 0.2 mm |
| Designed Current capacity | 2.5 kA @ 77K |
| Critical Current of cable | 3.6 kA @ 77K |
| Operational Current | 1 kA @ 77K |
| No. of HTS Tapes | 6 |
| No. of Copper tapes | 34 |
| $I = I_{op}/I_c$ | 0.7 |
| Upper limiting Current | 2.86 kA |
| Stekly Parameter ($\alpha$) | 0.736 |
| Energy Margin | 6.7 J/cc |

## IV. FABRICATION OF MAGNET

Figure.4 shows the schematic for the major components of the D-shaped magnet. The fabrication details of the winding pack in the cryogenic casing and vacuum casing are already reported [18]. The fabrication of HTS cable, winding pack, joints, terminations, instrumentation, and integration of D-shaped magnet are briefly discussed in the following sub-sections.

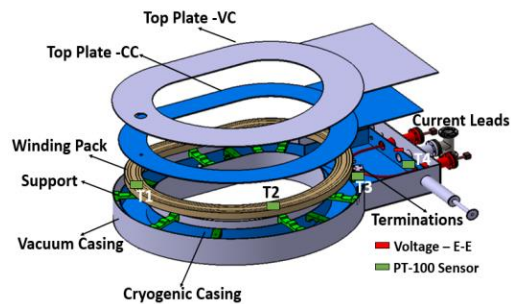

Fig.4. Major Components of D-shaped coil

### A. Fabrication of HTS Cable

HTS cable of length 35 m was fabricated using copper and REBCO tapes as mentioned in Table 2. The schema and fabrication processes of this cable are shown in the Fig.5. The individual HTS tape has the critical current of 600 A at 77 K @ self-field. In order to increase the copper fraction for stability during steady state operation, 34 numbers of 12 mm -width and 0.2 mm-thick copper tapes were used in this cable. During the cable fabrication, six numbers of stacks, each comprised of 5 copper + 1 HTS tapes, were prepared, ensuring their appropriate handling as shown in Fig.5a. These stacks were aligned with 2 pairs of top-down copper tapes as shown in Fig.5b.

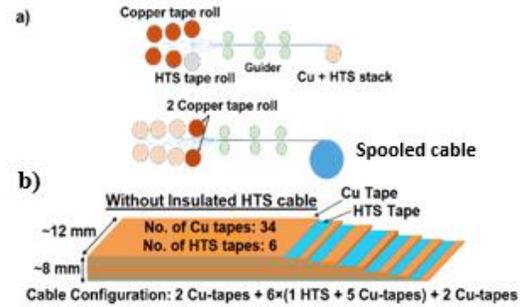

Fig.5. a) Process Schema and b) Architecture of stacked HTS cable

### B. Fabrication of HTS Winding Pack

The winding pack comprises two pancakes, each with six turns of length ~16 m stacked cable with a cross-section of ~13.5 mm × 9.5 mm including dry insulation of fiberglass tape. The winding setup consists of a former, a rotating base with jigs for cable feeding, and a riding spool. The bottom half of the cryogenic casing served as the former, defining the profile and dimensions of the winding pack for the D-shaped coil. The bottom pancake was wound in a clockwise direction, insulated with fiberglass tape, providing turn-to-turn isolation and maintaining mechanical integrity. Upon completion, the cable was re-fed from the riding spool to wind on the top pancake in an anti-clockwise direction. Figure.6b shows the winding pack of the D-shaped coil. After winding, both ends of the HTS cable were terminated in copper blocks via soldered joints as described in the subsequent section.

### C. Terminations with Winding Pack

Winding pack terminations of size 70 mm × 70 mm were prepared by soldering cable ends using the Pb-Sn wire with a melting point of about 183°C. The overlap length between the HTS cable and copper block was around 70 mm. The individual layers of the HTS cable were connected with Pb-Sn solder strips to form a monolithic cable terminal. The heater temperature during soldering was kept below 200°C for a short duration (~5 min) to prevent the HTS conductors from degradation.

### D. Fabrication and Integration of Current Leads

The current leads for interfacing the room temperature power supply with coil terminals at 77 K were designed using conventional copper materials. It consists of warm end for power supply, copper rod (Ø 10 mm and of 173 mm in length), flexible copper rope (Ø 10 mm and of 250 mm in length), and end connector for the HTS cable as shown in Fig. 6c and Fig.6d. The current lead assembly mounted on the 40 CF was leak tested under vacuum, and its insulation quality was verified with a Megger test at 1 kV. The insulation resistance of more than 500 GΩ was observed. The integration of the current leads assembly was carried out on a dedicated port of the cryogenic casing. The flexible copper rope was used for connecting the copper rod of the current lead with the winding pack via terminations, as shown in Fig. 6c.

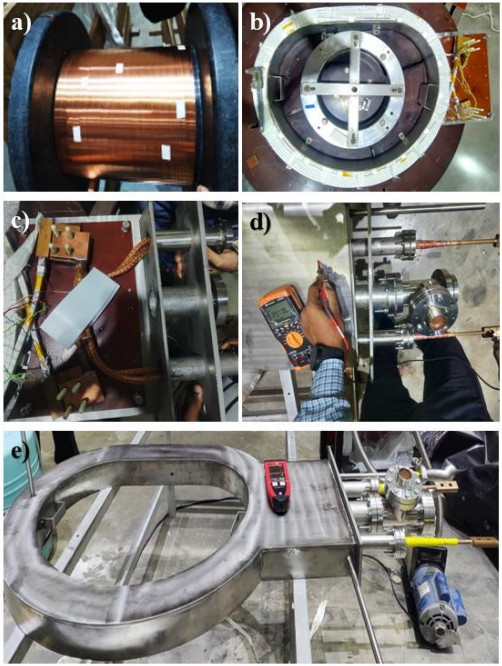

Fig. 6.a) HTS cable, b) Winding pack, c) Terminations with Cu blocks d) Integration of Current leads e) Cryogenic casing with winding pack

*E. Cryogenic casing of Winding Pack*

Figure.6e shows the cryogenic casing after enclosing the winding pack. After winding pack terminations, current leads integration and sensor mounting, Teflon supports were installed to mechanically secure the winding pack. The top L-section of the cryogenic casing was then positioned and tack-welded, followed by full TIG welding in a sequenced approach to minimize distortion. Upon welding completion, all external interface flanges & tubing were welded, and room-temperature electrical and mechanical inspections were performed. The cryogenic casing was then evacuated to $5 \times 10^{-2}$ mbar using a rotary pump. Leak testing was then performed with a Mass Spectrometer Leak Detector (MSLD), both before and after $LN_2$ thermal shock. No leaks were detected with a background sensitivity of $3.0 \times 10^{-9}$ mbar l/s for MSLD.

*F. Integration of Vacuum Casing and Leak Testing*

The final stage of fabrication involved assembling the cryogenic casing into the vacuum enclosure. The cryogenic assembly was placed into the bottom L-section of the vacuum casing. The top L-section was tack-welded, followed by full welding using a sequenced approach to reduce deformation. The evacuation and leak testing of the integrated D-shaped magnet was then further carried out under vacuum mode. No leak was detected with MSLD for background sensitivity of $3.5 \times 10^{-9}$ mbar l/sec.

*G. Sensors, Instrumentations, and DAQ*

Four PT-100 temperature sensors were strategically installed on the D-shaped coil to monitor critical temperatures as well as to provide information about the $LN_2$ level. PT-100 sensors were calibrated and tested at room temperature (RT) and at $LN_2$ temperature before installation. Winding pack end-to-end terminal (E-E) voltage tap were mounted to monitor the voltage characteristics during the testing. Sensors and voltage taps wires were anchored at the winding pack, terminations, and routed for their termination in feedthrough on the top of the coil. Voltage taps were continuously monitored using a high-sensitivity Nano-voltmeter to capture ultra-low signal variations during current ramping, steady-state, and discharging events. A high-speed modular PXI-based Data Acquisition System [19] was used to acquire sensor signals during the coil testing.

V. EXPERIMENTAL SETUP

Figure.7 shows the schematic and experimental setup for the testing of the D-shaped magnet. It consists of a vacuum system including a rotary pump with an isolation valve and gauges. The cryogenic circuit included a self-pressurized $LN_2$ Dewar, a long stem manual flow control valve at the inlet, and a vent valve at the outlet. The instrumentation feedthrough was installed at the 63 CF port. The controlled charging of the magnet was carried out after cooling down to 77 K using a programmable DC power supply. The magnetic field at various locations of the magnet was measured by the portable Lakeshore Tesla meter with a Hall probe. The data from all sensors, voltage taps, and output signals from the power supplies were acquired in the high-speed, modular PXI-based Data Acquisition System (DAQ) as mentioned earlier.

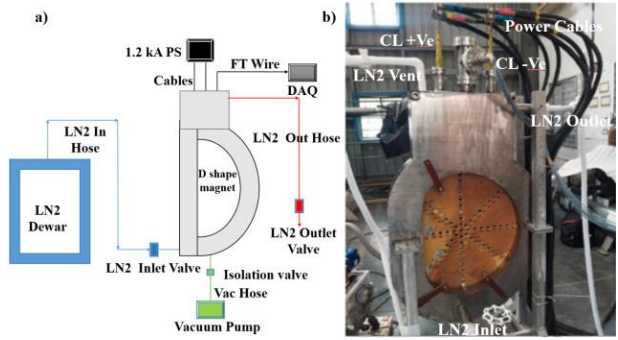

Fig.7. a) Schematic of experimental setup b) Test setup for D shaped magnet

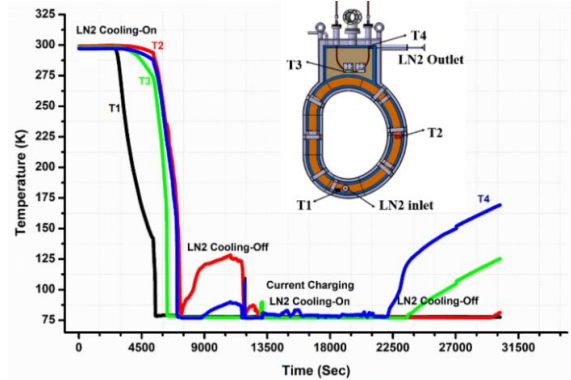

Fig.8. Cooldown profile of the D-Shaped magnet

## VI. RESULTS & DISCUSSION

The end-to-end terminal voltage of the D-shaped magnet was measured to be ~0.7 mV with a 100 mA current source at room temperature. The evacuation of the vacuum casing of the HTS magnet was carried out by a rotary pump, achieving a pressure up to $5.0 \times 10^{-2}$ mbar at room temperature within ~1.5 Hrs. $LN_2$ was then transferred to a cryogenic casing for the cooling of the HTS winding pack using a self-pressurized Dewar at 0.3 bar (g) via an inlet at the bottom of the magnet. The temperature and $LN_2$ level were monitored with temperature sensors mounted on the winding pack and on the cryogenic casing. The cooldown trends for the magnet using temperature sensors are shown in Fig. 8. The average cooldown rate was 1.45 K/min during the initial filling of the cryogenic casing. Further, the level of $LN_2$ was maintained by dynamic filling from a Storage Dewar of 230 L, maintained at 0.3-0.5 bar (g) using the control valve. T1 sensor (mounted at the bottom of the winding pack), submerged earliest, recorded a rapid temperature drop with an average cooling rate of ~3.71 K/min, indicating its direct immersion in $LN_2$. T3 sensor (mounted on the joint termination block) attained the temperature of 77 K before T2 (at the mid-section of the winding pack) and T4 (at the top plate of the cryogenic casing) sensors, which can be attributed to direct interaction of these sensors with cold $LN_2$ vapors during filling. T2 and T4 sensors show a similar cooling rate of ~ 2.47 K/min during filling. After cooling the cold mass for more than three hours, the magnet was then charged up to 1220 A, using two precision DC power supplies, namely, Bruker make 10 V, 1000 A, and AMI make 12 V, 200 A, connected in parallel with a ramp rate of 10 A/sec. Figure.9 shows the voltage characteristics of the D-shaped magnet up to 1.22 kA with a ramp rate of 10 A/s. The current operational time was ~450 s, including ramp-up and ramp-down.

The voltage drop of ~2.48 mV was observed with an inductance of 248 µH during the charging of the coil. The magnetic field was measured from the straight section inner edge to the curve section inner edge of the magnet at an interval of 35 mm. Figure 10 shows the comparison between the measured and estimated magnetic fields at different locations along the equatorial line in the bore of the magnet. The magnetic fields were 475 G -275 G @ 1.22 kA, at measurement locations, and are within 5% of the estimated values.

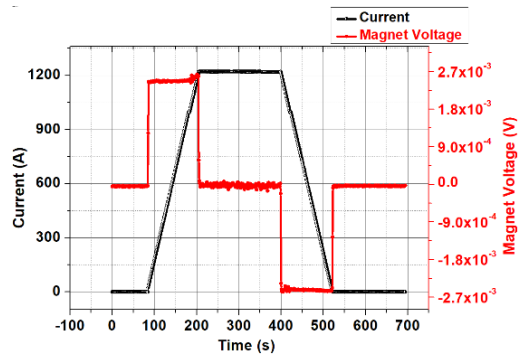

Fig.9. Current Vs voltage characteristics of HTS winding pack

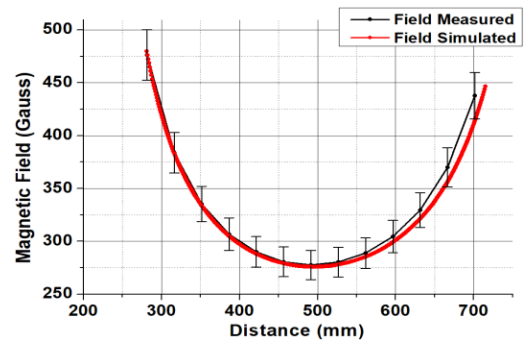

Fig.10. Comparison between measured and simulated field values

## VII. SUMMARY

The technical assessment and feasibility for the fabrication of a prototype D-shaped HTS coil was carried out to identify the challenges and requirements. The fabrication processes of long-length stacked HTS cable was realized using REBCO and copper tapes. The fabrication of the winding pack using HTS cable in the D-shaped profile was carried out using the cryogenic casing as the former. The fabrication and assembly of the various components of the magnet, such as casing, current leads, and joint termination blocks were meticulously carried out. The winding pack was successfully encased in a cryogenic casing, which was further enclosed in a vacuum enclosure. The magnet was successfully charged up to 1.22 kA at self-field. The simulated magnetic field profile for the prototype D-shaped magnet was validated with measured values and found within the error bar of 5%. The lessons learned and experience gained during the prototyping activities of this magnet will be helpful for the development of large-scale HTS magnets for tokamak applications.

## ACKNOWLEDGMENT

We would like to thanks colleagues from the IPR workshop, VESD, Mr. Kedar Bhope, and Mr.Mayur Mehta for their contribution in fabrication, leak testing, and DPT for this magnet.